\documentclass[aps,prb,showpacs,twocolumn]{revtex4}
\usepackage{graphicx}
\begin{document}

\title{Tight-binding parametrization for the chromium nitride: A NMTO study}

\author{Xiaofeng Wang}
\author{Min Zhu}
\author{Juan Huang}
\author{Wenhui Xie}\thanks{Electronic mail: whxie@phy.ecnu.edu.cn}

\affiliation{Engineering Research Center for Nanophotonics and Advanced Instrument, Department of Physics, East China Normal University, Shanghai, 200062, China}

\date{\today }

\begin{abstract}
We investigate the band structure of chromium nitride using the $N$th-order muffin-tin orbital (NMTO) based downfolding technique. The effective hopping Hamiltonian parameters are obtained using NMTO downfolded basis sets, which consist of Cr $d$ bands including $e_g$ and $t_{2g}$ states for both cubic and orthorhombic lattice. We analyze the chemical bonding and tight-binding parameters from the tight-binding Hamiltonian, further the effect of lattice distortion is discussed according to these parameters.
\end{abstract}
 \maketitle

\section{Introduction}

Chromium nitride (CrN) has great potential for industrial applications as hard, wear-, and corrosion-resistant coatings. It also attractes much attention due to its interesting fundamental physical properties and nonuniform picture of the electronic structure \cite{exp2009nmat,exp2011prbgall,exp2011prlshin,Gall}. The magnetic and structural properties of CrN have been well defined: at room temperature CrN is paramagnetic (PM) in the rock-salt (RS) structure (Space group Fm3m), but below N\`{e}el temperature $T_{N}$$\sim$286K it becomes an antiferromagnetism(AFM) orthorhombic phase (Space group Pnma)\cite{Eddin,Brwne,Tsuchiya,Constantin,Quintela}. The AFM magnetic ordering consists of ferromagnetic(FM) planes, stacked antiferromagnetically along the [110] direction every two FM Cr layers (AFM$^2_{[110]}$), which has been identified by neutron scattering\cite{Corliss}. In contrast to the general agreement on the magnetic ordering and crystal structure across $T_N$, the reports on its electronic properties are quite contradictory, with (i) metal to metal\cite{Brwne,Tsuchiya}, (ii) insulator to insulator\cite{Subramanya,Gall}, and also (iii) insulator to metal transition\cite{Constantin}.
A recent study has identified a softening of bulk CrN under pressure due to a manifestation of a strong competition between different types of chemical bond that are found at a crossover from a localized to a molecular-orbital electronic transition\cite{exp2009nmat}. Moreover,  experiment also indicates that the N-vacancy concentration and crystalline defects strongly affect electron transport\cite{exp2011prbgall}. Therefore the stoichiometric CrN has been considered a cubic paramagnetic correlated insulator at room temperature, but an orthorhombic antiferromagnet metal below $T_{N}$\cite{exp2011prlshin,exp2011prbgall}.

However, on the theoretical side, First-principles density function theory (DFT) calculations\cite{dftFilippetti,dftpickett} predict metallic ground states for both cubic and orthorhombic lattice. The results show Cr 3$d$ partial density of states (DOS) at the Fermi level ($E_F$), but no gap observed, which is in contrast to the experimental measurement\cite{exp2011prlshin}. After considering the On-site Coulomb repulsion for Cr 3$d$ electrons, DFT+U calculations for the AFM phase could open a direct gap when U value is larger than 3 eV\cite{dftldau}. Moreover, in agreement with experimental observation, the DFT calculations\cite{dftFilippetti} have confirmed that the ground state of orthorhombic CrN is AFM$^2_{[110]}$. It also predicts the ground state of the cubic lattice is antiferromagetism in which every one FM Cr layer stackes antiferromagnetically along the [110] direction (AFM$^1_{[110]}$). However, in contrast to the theoretical results, the experiments show that cubic CrN remains a paramagnetic insulator over the entire measured temperature range of 10-295 K\cite{Gall}.  Therefore, the relationship between magnetism and structure as well as the related electronic structure of CrN is still a controversial issue. In this paper, we employ the recently developed $N$th-order muffin-tin orbital (NMTO) method\cite{lmto1,lmto2,nmto} to produce real space Hamiltonian parameters for both cubic and orthorhombic phases, furthermore investigate the hopping integrals of the tight-binding Cr $d$ bands and analyze the chemical bonding.

\section{The crystal structures and computational method}

The two kinds of structures are very well known and are described as bellow: One is the cubic rock-salt structure with a lattice constant of $a$ = 4.148 {\AA}\cite{exp2009nmat}. Each Cr atom is surrounded by six N atoms, thereby providing the octahedral environment at the Cr site, which leads to the splitting of the degenerate $d$ orbital into $t_{2g}$ and $e_{g}$ states. Every Cr atom has 12 first nearest neighbor (first-NN) Cr atoms and 6 second nearest neighbor (second-NN) Cr atoms.
Below the N\`{e}el temperature, CrN has the orthorhombic antiferromagnetic structure (lattice constants are $a$ = 5.757 {\AA}, $b$ = 2.964 {\AA}, $c$ = 4.134 {\AA})\cite{Corliss}, in which the lattice expands about 1.8\% along [1$\bar{1}$0] direction and shrinks about -1.1\% along [110] direction of the original cubic structure, respectively. Since there is the simultaneous occurrence of both magnetic order and structure transition, in order to separate the effect of magnetism and structure distortion, we ignore the magnetic ordering for the orthorhombic phase. In the orthorhombic lattice, each Cr atom is still sixfold coordinated by nitrogen atoms and the corresponding octahedra is distorted due to the tiny shear deformation in $xy$ plane, while $c$-axis is roughly unchanged.
The 12 first-NN Cr-Cr pairs are split into two longer and two shorter pairs in $xy$ plane and roughly unchanged 8 Cr-Cr pairs out of $xy$ plane. It is also noted that the bond distance of Cr-N as well as second-NN Cr-Cr pairs remain nearly constant.

In the present study we firstly use the tight-binding linear muffin-tin orbital (TB-LMTO) method with the atomic-sphere approximation technique\cite{lmto1,lmto2} to generate electronic structure. The non-spin-polarized calculations for both the cubic and orthorhombic CrN phases have been performed within Local Density Approximation (LDA). Then NMTO method is used to carry out the localized Wannier function which is the orthogonalized L\"{o}wdin functions from the atomic orbitals. The downfolding technique in the NMTO method allows to produce minimal bands which follows exactly the bands derived with the large basis set, which actually is the maximum localized Wannier function. We choose all of Cr $d$ orbitals to form the minimal basis set, and further the corresponding hopping integrals of the tight-binding Cr $d$ bands are derived from the NMTO calculations.

\section{Results and discussions}\label{ss2}
\subsection{LDA band structure}\label{ss2}
The non-spin-polarized band structures of CrN are calculated with the full atomic orbitals, which are shown in the left-panels of Fig.1 and Fig.2 for the cubic lattice and the orthorhombic lattice, respectively. In the cubic lattice, the high-symmetry points in Billouin Zone are: $L=(0.5,0.5,0.5)$, $G=(0,0,0)$, $X=(0,1,0)$, $W=(0.5,1,0)$, $L=(0.5,0.5,0.5)$,$K=(0,0.75,0.75)$, $G=(0,0,0)$ and $Z=(0,0,1)$. In the orthorhombic lattice, the high symmetry points in Billouin Zone are: $L=(0.5,0,0.5)$, $G=(0,0,0)$, $X=(0.5,0.5,0)$, $W=(0.75,0.25,0)$, $L=(0.5,0,0.5)$, $K=(0.375,0.375,0.75)$, $G=(0,0,0)$, and $Z=(0,0,1)$. The chosen high symmetry lines would pass through similar way in Billouin Zone of both the cubic and orthorhombic lattices, so that we could compare the discrepancy of band structures between two crystal structures.

The electronic structure of CrN is extensively investigated previously. The strong hybridization has been found between Cr $d$ and N $p$ states. The wide band indicates the strong $p$-$d$ bonding between Cr $d$ orbitals and N $p$ orbitals. Comparing the bands shown in Fig.1 and Fig.2, it is found that the band structures of both phases are quite similar except that the small band splitting about 10 meV induced by orthorhombic distortion. As shown in Fig.1, the lowest three bands are primarily $p$ states of N distributing from -9 eV to -2 eV. The $t_{2g}$-derived states at the Cr site span the energy of about -3.9 eV$\sim$0.9 eV, while the $e_{g}$-derived states of Cr occupy the energy range from -0.7 eV  to 4.2 eV. Around the Fermi level, the bands are primarily Cr 3$d$ states of $t_{2g}$ manifold, with $e_{g}$ bands situating in higher energy due to the crystal-field splitting. In the cubic lattice, the $d$-orbitals are simply split to $t_{2g}$ ($d_{xy}$, $d_{yz}$, $d_{xz}$) and $e_g$ ($d_{3z^2-1}$, $d_{x^2-y^2}$) mainfolds due to symmetry of the CrN$_6$ octahedra. While in the orthorhombic lattice, as the shear deformation in $xy$ plane reduces the symmetry of distorted octahedra, $t_{2g}$ is further split to $d_{xy}$, $d_{yz}$ and $d_{xz}$ states, and $e_g$ is split to $d_{3z^2-1}$ and $d_{x^2-y^2}$ states, respectively.


\subsection{Downfolding onto the Cr-$d$ manifolds for the cubic lattice}

Since the band gap observed experimentally in paramagnetic cubic phase could not be captured by LDA or GGA calculations, it suggestes that CrN should be electron correlated system\cite{exp2011prlshin}. Therefore, constructing a model hamiltonian which could be used in many body picture to perform the realistic calculation is highly expected. We propose a downfolding scheme in which all orbitals of whole atoms except Cr-$d$ are downfolded. Both Cr $t_{2g}$  and $e_{g}$ states are considered to constitute the effective orbitals because electrons occupation on higher e$_g$ state could not be ignored in both cubic and distorted structures. For the local coordination, the $x$ and $y$ coordinate axis of CrN$_6$ octahedra are chosen in original direction of the coordinate axis of the cubic lattice. The effective hopping Hamiltonian matrix of dimension $5\times5$, built up by the five $d$ effective orbitals. Fourier transformation of the downfolded Hamiltonian (H(k)$\rightarrow$H(R)) gives a TB electronic Hamiltonian in real space consisting of hopping over up to five NNs.  For many studies, it is desirable to have a shorter-range Hamiltonian which can be achieved by the downfolding technique. Taking the first nearest neighbors and the second nearest neighbors into account, the total Hamiltonian matrix can be expressed as follow:
\begin{equation}
H=t^{000}_{m',m}+\sum\limits_{n=1}^{12}t^{\vec {R_n}}_{m',m} \times e^{i\vec {K} \cdot \vec {R_n}}+
\sum\limits_{n=1}^{6}t^{\vec {R^{'}_n}}_{m',m} \times e^{i\vec {K} \cdot \vec {R^{'}_n}}
\end{equation}
In this formula, $R_{n}$ and $R^{'}_n$ are the coordinates of 12 first nearest neighbors and 6 second neighbors, respectively. Here only one representative hopping matrix at each NN is shown as bellow, but due to the crystal symmetry all the other NN hopping integrals can be derived from proper unitary transformation.

The basis set of Cr-$d$ NMTO orbitals:
\begin{equation}|\chi^{\bot}\rangle=\left\{|xy\rangle,|yz\rangle,|zx\rangle,|3z^{2}-1\rangle,|x^{2}-y^{2}\rangle\right\}\end{equation}

The on-site hopping integral (The unit is eV and $E_{F}$ = 3.2685eV.):
\begin{equation}t^{000}_{m',m}=\left( {\begin{array}{*{20}{c}}
3.1618 &0 &0 &0 &0 \\
0 &3.1618 &0 &0 &0 \\
0 &0 &3.1618 &0 &0 \\
0 &0 &0 &5.3497 &0 \\
0 &0 &0 &0 &5.3497 \\
\end{array}} \right)
\end{equation}

The first nearest neighbor (The unit is meV):
\begin{equation}t^{\frac{1}{2}\frac{1}{2}0}_{m',m}=\left( {\begin{array}{*{20}{c}}
-354&      0&      0&     -219&      0\\
 0&        88&    -92&     0&        0\\
 0&       -92&     88&     0&        0\\
-219&      0&      0&      21&       0\\
 0&        0&      0&      0&     -168\\
\end{array}} \right)
\end{equation}

The second nearest neighbor (The unit is meV):
\begin{equation}t^{001}_{m',m}=\left( {\begin{array}{*{20}{c}}
 10&      0&      0&      0&      0\\
 0&     -177&     0&      0&      0\\
 0&      0&     -177&     0&      0\\
 0&      0&      0&     -530&     0\\
 0&      0&      0&      0&       33\\
\end{array}} \right)
\end{equation}

The on-site term $t^{000}_{m',m}$ is diagonal, and doubly degenerated $e_{g}$ and triply $t_{2g}$ are presented. The crystal-filed splitting between $e_{g}$ and $t_{2g}$ state is 2.2 eV. In the downfolding procedure, we have obtained the hopping matrix elements of all Cr-$d$ orbitals, thus we can derive the energy bands from these hopping matrixes throughout the Brillouin zone. Hopping integrals further than third NN are small. The largest values are 35 meV for third NN,  59 meV for fourth NN and 54 meV for fifth NN, respectively. The hopping of third NN should be pass through chains like Cr[000]$\cdot$$\cdot$$\cdot$Cr[$\frac{1}{2}\frac{1}{2}0$]$\cdot$$\cdot$$\cdot$Cr[110] and the hopping of forth-NN should be pass through chains like Cr[000]$\cdot$$\cdot$$\cdot$Cr[$\frac{1}{2}\frac{1}{2}0$]$\cdot$$\cdot$$\cdot$Cr[1$\frac{1}{2}\frac{1}{2}$], so that they are small. The others of matrix elements are mostly zero. Further NN hopping integrals are negligible.

From the effective TB second-NN Hamiltonian, the best optimized downfolding bands are plotted with black dash lines in the right panel of Fig.1. The bands obtained by downfolding all the other channels expect the Cr-d channels include only the first-NN and second-NN Hamiltonian in real space. There is distinguish discrepancy between the full bands and downfolding bands, which indicates that the long-range hopping have certain contribution. We find that the third NN and fourth NN terms are necessary to fit the full bands.

Furthermore, the second-NN Hamiltonian derived from downfolded bands are expressed in terms of the Slater-Koster integrals $dd\sigma$, $dd\pi$, $dd\delta$, which could be used to understand the chemical bonding between Cr atoms on different sites. Starting from the first-NN hopping parameters of $t_{2g}$, we obtain $dd\sigma$=532meV, $dd\pi$=4meV, and $dd\delta$=180meV, while from the first-NN hopping parameters of $e_{g}$ we obtain $dd\pi$=168meV and $\frac{1}{4}$$dd\sigma$+$\frac{3}{4}$$dd\delta$=21meV. The difference of chemical bonding obtained between $t_{2g}$ and $e_g$ orbitals is due to the influence of N-2$p$ obitals. Because the $e_g$ orbitals have strong $pd\sigma$ antibonding with N-2$p$ states, while $t_{2g}$ orbitals form weak antibonding $pd\pi$ coupling with N-2$p$ states. Since the influence of N-2$p$ tail is captured in the NMTO downfolding procedure, therefore the two-center integral treatment of Slater-Koster integrals formation provide different bonding strength. Moreover, from the second-NN hopping parameters, we obtain that the chemical bonding are $dd\pi$=-117meV, $dd\delta$=9.8meV for $t_{2g}$ orbital and $dd\sigma$=-530meV, $dd\delta$=32.7meV for  $e_g$ orbital, respectively. The chemical bonding between $e_g$ orbitals are dominated by $dd\sigma$ plus a little $dd\delta$ without $dd\pi$, while between $t_{2g}$ orbitals are $dd\pi$ dominated.

It could be found that the effective bonding connected to second-NN Cr atoms are actually $indirect$ bonding, while for first-NN Cr atoms are mainly $direct$ bonding between Cr $d$ orbitals, if we perform downfolding of the Hamiltonian in the larger basis including N 2$p$ orbitals. The second-NN $indirect$ bonding between $t_{2g}$ orbitals are primarily dominated by the nitrogen-mediated $pd\pi$ coupling, while bonding between $e_{g}$ orbitals are primarily dominated by the nitrogen-mediated $pd\sigma$ coupling, along $x$, $y$ and $z$-axis directions via the Cr$\cdot$$\cdot$$\cdot$O$\cdot$$\cdot$$\cdot$Cr chain. As the $pd\sigma$ and $pd\pi$ bonding are very strong ($\sim$ 1.5 eV), so that the $indirect$ bonding ($\sim$ $\frac{t^{2}_{pd}}{\varepsilon-\varepsilon_{p}}$) of second-NN has comparable strength with the $direct$ $dd$ bonding of first-NN.

\begin{figure}
\centering
\includegraphics[width=8cm]{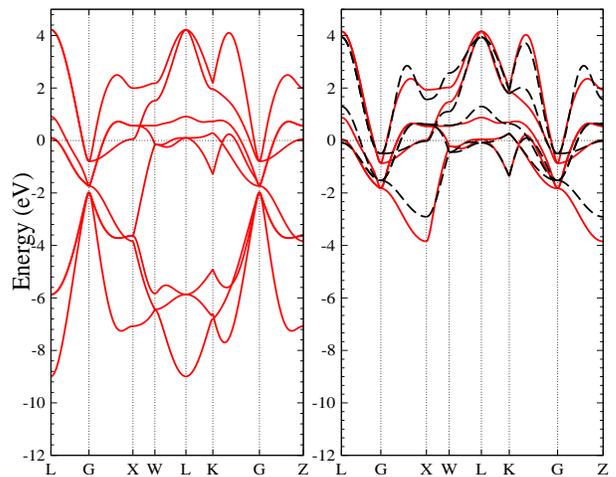}
\caption{(Color online) The band structure calculated with full basis (left panel) and the downfolded Cr-d bands (right panel) of cubic CrN.
In the right panel also shows the best-fit TB bands with hopping integrals up to second-NN  (black dashed lines). The LDA $d$ bands are also shown for compasion (red solid lines).}
\end{figure}

\subsection{Downfolding onto the Cr-$d$ manifolds for the orthorhombic lattice}

We now turn to the orthorhombic lattice. As same as the cubic case, the effective TB Hamiltonian is of dimension 5$\times$5, which is defined on the basis of the effective $t_{2g}$ and $e_g$ orbitals for the Cr sites of the orthorhombic lattice. The hopping integrals of Cr-$d$ Wannier orbitals in the real space are presented up to the second NN. Since the Cr atoms centered CrN$_6$ octahedra is slightly distorted by the orthorhombic distortion, the 12 first-NN terms in Hamiltonian are split into three groups according to crystal symmetry: two hopping integrals in $xy$ plane connect to ($\frac{1}{2}\frac{1}{2}0$) and (-$\frac{1}{2}$-$\frac{1}{2}0$) sites, other two hopping integrals in $xy$ plane connect to ($\frac{1}{2}$-$\frac{1}{2}0$) and (-$\frac{1}{2}\frac{1}{2}0$) sites, and the rest eight hopping integrals are corresponding to the sites out of $xy$ plane. Similar with cubic case, only one representative hopping matrix for each group is shown in the following.

\begin{figure}
\centering
\includegraphics[width=8cm]{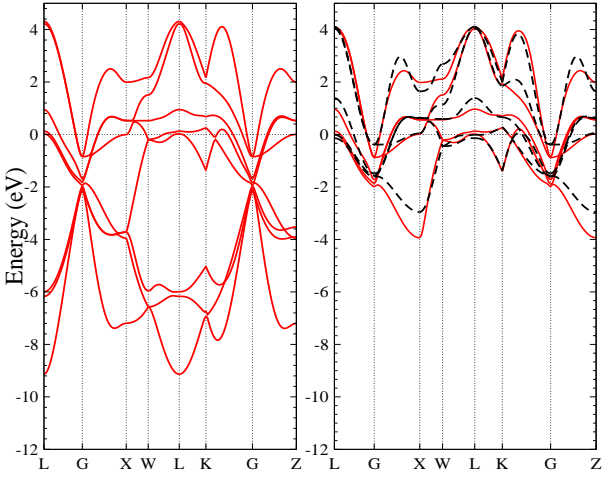}
\caption{(Color online) The band structures calculated with full basis (left panel) and the downfolded Cr-d bands (right panel) of orthorhombic CrN (Pnma). The right panel show the best-fit TB bands with hopping integrals up to second-NN (black dashed lines). The LDA $d$ bands are also shown for compasion (red solid lines).}
\end{figure}

For convenient, the $a$ and $b$ axis are further rotated 45$^\circ$ in NMTO calculations, pointed to (110) and (1$\bar{1}$0) of the orthorhombic lattice, while the $c$ axis is unchanged. In this local coordination system, the $a$, $b$ and $c$ axis are as same as that in the cubic structure, thus we can compare the hopping integral directly.

The basis set of Cr-$d$ NMTO orbitals:
\begin{equation}|\chi^{\bot}\rangle=\left\{|xy\rangle,|yz\rangle,|zx\rangle,|3z^{2}-1\rangle,|x^{2}-y^{2}\rangle\right\}\end{equation}

The on-site term (The unit is eV and $E_{F}$ = 3.4286 eV.):

\begin{equation}t^{000}_{m',m}=\left( {\begin{array}{*{20}{c}}
3.3194   &0       &0      &0.0353  &0 \\
0      &3.3186  &-0.008 &0       &0 \\
0      &-0.008  &3.3186 &0       &0 \\
0.0353 &0       &0      &5.5598  &0 \\
0      &0       &0      &0       &5.5641 \\
\end{array}} \right)
\end{equation}

The first nearest neighbor (The unit is meV):

\begin{equation}t^{\frac{1}{2}\frac{1}{2}0}_{m',m}=\left( {\begin{array}{*{20}{c}}
-416    &0       &0       &-189    &0 \\
0       &101     &-81     &0       &0 \\
0       &-81     &101     &0       &0 \\
-189    &0       &0       &26      &0 \\
0       &0       &0       &0       &-192 \\
\end{array}} \right)
\end{equation}

\begin{equation}t^{\frac{1}{2}-\frac{1}{2}0}_{m',m}=\left( {\begin{array}{*{20}{c}}
-304&         0&     0&      254&    0\\
 0&          76&     104&    0&      0\\
 0&          104&    76&     0&      0\\
 254&        0&      0&      15&     0\\
 0&          0&      0&      0&     -159\\
\end{array}} \right)
\end{equation}

\begin{equation}t^{0\frac{1}{2}\frac{1}{2}}_{m',m}=\left( {\begin{array}{*{20}{c}}
 87&       -2&      -93&     -5&     2\\
 -2&       -360&     4&     112&    -193\\
 -93&       4&       88&     0&      4\\
-5&        112&      0&     -127&    -86\\
2&        -193&      4&     -86&     -28\\
\end{array}} \right)
\end{equation}

The second nearest neighbor (The unit is meV):
\begin{equation}t^{001}_{m',m}=\left( {\begin{array}{*{20}{c}}
 9&      0&       0&      -7&      0\\
 0&     -179&    -2&       0&      0\\
 0&     -2&     -179&      0&      0\\
-7&      0&       0&     -559&     0\\
 0&      0&       0&       0&      27\\
\end{array}} \right)
\end{equation}

The best optimized downfolded bands are also plotted in the right-hand panel of Fig.2 (black dash lines). Similar with that of the cubic lattice, we could find the discrepancy between the full bands and downfolded bands, because the bands obtained by downfolding only take account of the first-NN and second-NN Hamiltonian in real space.

It is found that the on-site terms are not diagonal unless in \{ 0.9998$d_{xy}$+0.0158$d_{3z^{2}-1}$, $\frac{\sqrt{2}}{2}$$d_{xz}$+$\frac{\sqrt{2}}{2}$$d_{yz}$, $\frac{\sqrt{2}}{2}$$d_{xz}$-$\frac{\sqrt{2}}{2}$$d_{yz}$, -0.0158$d_{xy}$+0.9998$d_{3z^{2}-1}$, $d_{x^{2}-y^{2}}$ \} representation, corresponding to eigenvalue \{3.3106, 3.3188, 3.3266, 5.5604, 5.5641\}, respectively. Since the orthorhombic distortion in the $xy$ plane, the $d_{xz}$ and $d_{yz}$ orbitals are not orthogonal, while the orbital combinations $\frac{\sqrt{2}}{2}$$d_{xz}$+$\frac{\sqrt{2}}{2}$$d_{yz}$ and $\frac{\sqrt{2}}{2}$$d_{xz}$-$\frac{\sqrt{2}}{2}$$d_{yz}$ are orthogonal, whose directions point to the $a$ and $b$ axis of the orthorhombic lattice, respectively. The orbital combinations of 0.9998$d_{xy}$+0.0158$d_{3z^{2}-1}$ and -0.0158$d_{xy}$+0.9998$d_{3z^{2}-1}$ indicate that the pure state of $d_{xy}$ or $d_{3z^{2}-1}$ orbital in the cubic lattice slightly mix each other due to the distortion effect induced by the expansion of $a$ axis and shrinking of $b$ axis in the orthorhombic lattice.

The influence of the orthorhombic distortion on the hopping integrals of first-NN and second-NN is clearly indicated, when we compare the hopping terms $t^{\frac{1}{2}\frac{1}{2}0}_{m',m}$, $t^{\frac{1}{2}-\frac{1}{2}0}_{m',m}$, $t^{0\frac{1}{2}\frac{1}{2}}_{m',m}$ and $t^{001}_{m',m}$ of the distorted lattice with the corresponding hopping terms of the cubic lattice. It indicates that most of first-NN hopping terms have a approximated relationship: $t^{NN}_{m',m}$ (orthorhombic) + $t^{NN}_{m',m}$(orthorhombic) $\simeq$ 2 $\times$ $t^{NN}_{m',m}$ (cubic), while second-NN hopping matrix $t^{001}_{m',m}$ change a little. Thus, the effect of orthorhombic distortion on electronic structure is almost offset in these terms. The net contributions primarily originate from the first-NN hopping terms of $t_{xz,yz}$ and $t_{3z^2-1,xy}$ between the Cr atoms in $xy$ plane, which also indicate the nature of the orthorhombic distorted effect similar with that found in on-site terms.
Therefore, the total energy should change very tiny when the lattice of CrN distorts from the original cubic to the orthorhombic structure, as been shown in the total energy results of the previous nonmagnetic first-principles electronic structure calculations \cite{dftFilippetti}.
\section{Summary}\label{ss4}

In conclusion, by using the downfolding technique with the NMTO method, we have investigated the electronic structure of both the cubic and orthorhombic structure CrN, and obtained the parameters for the TB Hamiltonian with downfolded Cr $d$ orbitals. Further the chemical bonding and the distortional effect are analyzed according to the TB parameters. It is found that the first-NN hopping integral $t_{xz,yz}$ and $t_{3z^2-1,xy}$ should be responsible for the effect of orthorhombic distortion and long range hopping terms have little contribution. Therefore, the non-spin-polarized result with Cr $d$ orbitals involving second-NN hopping terms might be a good starting point for the many-body calculations to consider the complex physical properties.

\section{ACKNOWLEDGEMENT}

This work is supported  by Nature Science Foundation of China (Grant No. 10704024), by Shanghai Rising-Star Program (Grant No. 08QA14026), and by Shanghai Scientific Project
(Grant No. 08JC1408400).


\begin{thebibliography}{99}

\bibitem{exp2009nmat} F. Rivadulla, M. B. Nobre L¨®pez, C. X. Quintela, A. P. Neiro, V. Pardo, D. Baldomir, M. A. L¨®pez-Quintela, J. Rivas, C. A. Ramos, H. Salva, J. S. Zhou, and J. B. Goodenough, Nature Materials \textbf{8}, 947-951 (2009).

\bibitem{exp2011prbgall} X. Y. Zhang, J. S. Chawla, B. M. Howe, and D. Gall,  Phys. Rev. B {\bf 83}, 165205 (2011).

\bibitem{Gall} D. Gall, C.-S. Shin, R. T. Haasch, I. Petrov, and J. E. Greene, J. Appl. Phys. \textbf{91}, 5882 (2002).

\bibitem{exp2011prlshin} P. A. Bhobe, A. Chainani, M. Taguchi, T. Takeuchi, R. Eguchi, M. Matsunami, K. Ishizaka, Y. Takata, M. Oura, Y. Senba, H. Ohashi, Y. Nishino, M. Yabashi, K. Tamasaku, T. Ishikawa, K. Takenaka, H. Takagi, and S. Shin, Phys. Rev. Lett.
\textbf{104}, 236404 (2010).

\bibitem{Eddin} M. N. Eddin, F. Sayetat and E. F. Bertaut, C. R. Seances Acad. Sci., Ser. B \textbf{269}, 574 (1969).

\bibitem{Brwne} J. D. Browne, P. R. Liddell, R. Street and T. Mills, Phys. Status solidi (a) \textbf{1}, 715 (1970).

\bibitem{Tsuchiya}Y. Tsuchiya, K. Kosuge, Y. Ikeda, T. Shigematsu, S. Yamaguchi, and N. Nakayama, Mater. Trans., JIM \textbf{37}, 121 (1996).

\bibitem{Constantin} C. Constantin, M. B. Haider, D. Ingram, and A. R. Smith, Appl. Phys. Lett. \textbf{85}, 6371 (2004).

\bibitem{Quintela} C. X. Quintela, F. Rivadulla, and J. Rivas, Appl. Phys. Lett. \textbf{94}, 152103 (2009).

\bibitem{Corliss} L. M. Corliss, N. Elliott, and J. M. Hastings,
Phys. Rev. \textbf{117}, 929 (1960).

\bibitem{Subramanya} P. Subramanya Herle, M. S. Hedge, N. Y. Vasathacharya, and S. Philip, J. Solid State Chem. \textbf{134}, 120 (1997).

\bibitem{dftFilippetti} A. Filippetti, and N. A. Hill,
Phys. Rev. Lett. \textbf{85}, 5166 (2000).

\bibitem{dftpickett} A. Filippetti, W. E. Pickett, and B. M. Klein,
Phys. Rev. B \textbf{59}, 7043 (1999).

\bibitem{dftldau} A. Herwadkar and W. R. L. Lambrecht, Phys. Rev. B \textbf{79}, 035125 (2009).

\bibitem{lmto1} O. K. Andersen, Phys. Rev. B  \textbf{12}, 3060 (1975).

\bibitem{lmto2} O. K. Andersen, and O. Jepsen, Phys. Rev. Lett. \textbf{53}, 2571 (1984).

\bibitem{nmto} O. K. Andersen, and T. Saha-Dasgupta, Phys. Rev. B \textbf{62}, R16219 (2000).







\end{thebibliography}
\end{document}